\documentclass{article}
\usepackage{psfig}
\usepackage{bioinformatics}

\newcommand{\be}{{\bf e}}
\newcommand{\s}{{\cal S}}
\newcommand{\p}{{\cal P}}
\newcommand{\D}{{\cal D}}
\newcommand{\G}{{\cal G}}
\begin{document}

\title{Algorithm for Finding Optimal Gene Sets in Microarray Prediction}

\author{J.M.Deutsch \\
University of California, Santa Cruz, U.S.A.}
%\noreportnumber
\maketitle    % End title section
\abstract
{
\noindent
{\bf\large Motivation:}
Microarray data has been recently been shown to be efficacious in
distinguishing  closely related cell types that often appear in the
diagnosis of cancer. It is useful to determine the minimum number of genes
needed to do such a diagnosis both for clinical use and to determine the
importance of specific genes for cancer.  Here a replication algorithm is
used for this purpose.  It evolves an ensemble of predictors, all using
different combinations of genes to generate a set of optimal predictors.

\noindent
{\bf\large Results:}
We apply this method to the leukemia data of the Whitehead/MIT group that
attempts to differentially diagnose two kinds of leukemia, and also to
data of Khan et. al. to distinguish four different kinds of childhood
cancers. In the latter case we were able to reduce the number of genes
needed from 96 down to 15, while at the same time being able to perfectly
classify all of their test data.

\noindent
{\bf\large Availability:}
http://stravinsky.ucsc.edu/josh/gesses/

\noindent
{\bf\large Contact:}
josh@physics.ucsc.edu
}

\newpage

\section*{Introduction}

cDNA microarrays have been used with great success to
distinguish cell types from each other, and hence has promising applications
to cancer diagnosis.  While the histopathology of two cells may appear
very similar, their clinical behavior, such as their response to drugs
can be drastically different. The use of microarrays has been shown in
many cases to provide clear differential diagnosis rivaling or surpassing
other methods and leads to a clustering of data into different
forms of a 
disease\cite{Derisi:cancer96,Alon:colon,Perou:breast,Zhu:megalo,Wang:ovarian,Schummer:ovarian,Zhang:normcanc,Alizadeh:lymphoma,Golub:leukemia,Khan:srbct}.

Many approaches have been used to classify microarray data. These
include the use of artificial neural
networks\cite{Khan:srbct,Furey:svm2000}, logistic regression\cite{LiYang:mingenes}, 
support vector machines\cite{Brown:svm,Furey:svm2000}, coupled two-way clustering\cite{Getz:twoway},
and weighted votes - neighborhood analysis\cite{Golub:leukemia}. For much of the data 
all these techniques appear to give similar
results and their performance improves as the amount and quality of data
increases. For example work on the classification
of two different leukemias\cite{Golub:leukemia} attempts to classify 34 test samples based
on 38 training samples. On prediction of test data, different predictors 
make anywhere from 0 to 5 mistakes. On the other hand, recent work on small
round blue cell tumors (SRBCT) attempted to classify 20 test samples based on
63 training samples\cite{Khan:srbct}. They were able to classify all test data correctly
into one of four separate categories.
They were able to do this with a single layer neural
network that considered only 96 genes.

To classify samples using microarray data, it is necessary to
decide which genes should be included in a predictor. Including too few genes will not discriminate
in a detailed enough manner to classify test data correctly. Having too many genes is
not optimal either, as many of the genes are largely irrelevant to the diagnosis
and mostly have the effect of adding noise, decreasing the 
``information criterion''\cite{LiYang:mingenes,akaike:info,Burnham:model,Schwarz:estimating}. 
This is particularly severe
with a noisy data set and few subjects. Therefore an effort is made
to choose an optimal set of genes for which to start the training of a predictor. This is
done in a variety of different ways, such as a kind of neighborhood analysis\cite{Golub:leukemia},
principle component analysis\cite{Khan:srbct}, or gene shaving\cite{Tibshirani:shaving}
A predictor can then be developed from this carefully chosen subset of genes.

Recent work\cite{LiYang:mingenes} addressed the problem of gene selection for
a leukemia data set\cite{Golub:leukemia}. They initially ranked genes 
as had been done in the first analysis\cite{Golub:leukemia} and used the top
ranked genes. They varied the number they included and found no clear indication of
any optimum number aside from the conclusion that the number should be much smaller
than the 50 that had been originally used\cite{Golub:leukemia}.

Here we develop gene selection further by making it an integral part of the
prediction algorithm itself. Instead of using all of the highest ranked
genes, we find an effective method to greatly reduce this number.
This can be done because gene expression tends to be highly correlated,
making many of the initially chosen genes redundant or even deleterious
because of the problem of added noise.

The method introduced here is
named GESSES (genetic evolution of sub-sets of expressed sequences).
It makes use of  a kind of evolutionary algorithm known
as a replication algorithm that has been extensively used
in quantum simulations \cite{ceperley} and protein folding\cite{Garel:repl}. 
It finds a set of highly relevant
genes by considering a whole ensemble of predictors, each of which use a
different sets of genes. As the predictors evolve, more genes are added
to each predictor. It eventually produces an ensemble of predictors each
of which can be tried on test data.  

In the case of
small round blue cell tumors, GESSES reduces the number of genes 
from 96 down to below 15 while still predicting the test data perfectly.

Aside from optimizing predictive capabilities, it is hoped that
GESSES will have applications in the clinical diagnosis
of cancer\cite{HeFriend:newsviews}. For this purpose it is important
to use as few genes as possible and still obtain an accurate diagnosis
of the disease. 

With the
same algorithms applied to leukemia data of Golub et. al., we find
conclusions in accord with Li and Yang\cite{LiYang:mingenes}
that there is no clear indication of an optimum number of genes
to use in a predictor. We find
a range of predictors some that predict the test data perfectly but many
predictors
that get several samples wrong. This is also in accord with other groups 
work\cite{Golub:leukemia,Furey:svm2000}. Without
further data or further biological information, it is probably not possible
to do better than this. 
This paper is organized as follows. We discuss the algorithm used
in detail by first defining the terminology and concepts used.
Then we discuss the predictor used, the kind of evolutionary algorithms
and the scoring function. We then apply this to two data sets,
the SRBCT and leukemia data. Last, we make some concluding remarks.

\section*{The Algorithm}
\subsection*{Terminology}
\label{sub:terminology}

We have samples of  microarray training data $\D_t \equiv \{D_1,D_2,\dots \}$ 
with each sample consisting 
of $N$ genes.
Corresponding to the $ith$ gene of a sample is its expression level $e_i$.
The complete set of genes $\G_t$ is the collection of genes $1$ through 
$N$ and we 
will consider subsets of ${\G}_t$, for example the subset
$\alpha_1,\alpha_2,\dots,\alpha_m$. (e.g. genes 2,5, and 9),
which we denote $G_\alpha$. For this subspace of genes 
the vector of expression levels 
$ \be_\alpha \equiv (e_{\alpha_1}, \dots, e_{\alpha_m})$ .
The number $m$ of genes in this subset is denoted $\vert \alpha\vert$, which
in this case is $m$.

Each sample $D$ has a classification of type $T$, in this case the type of cancer, which
can take one of $N_T$ values. The set of possible types is denoted $\cal T$.

We introduce the usual definition of the Euclidean distance between samples
$D_a$ and $D_b$ on the subspace $G_\alpha$:
\begin{equation}
\label{eq:distance}
{d^2}_\alpha (a,b) ~=~  {1\over \vert\alpha\vert}\vert {\be^a}_\alpha - {\be^b}_\alpha|^2 
~=~ {1\over \vert\alpha\vert}\sum_{i=1}^{\vert\alpha\vert} \vert{e^a}_{\alpha_i}-{e^b}_{\alpha_i}\vert^2
\end{equation}
where ${\be^a}_\alpha$ and ${\be^b}_\alpha$ are the expression levels of samples $a$ and $b$ 
respectively, for genes $\alpha_1,\alpha_2,\dots,\alpha_m$. 

\subsection*{Predictor}
\label{sub:predictor}

We define a predictor $\p$ as a function that takes a data sample $D$ and outputs
a type $T$, in this case the type of cancer that is associated with that data. That is
$\p(D) \rightarrow T$.

In this work we will use a k-nearest neighbor search\cite{Duda:pattern} to construct the predictor.
In the results reported below, we use $k=1$, that is,
the set of samples that forms the training data ${\cal D}_t$ are compared with the 
target sample $D$ by finding the distance using eqn. (\ref{eq:distance}) between
$D$ and each vector in the training set. The sample in the training set
closest to $D$ gives the classification $T$ of $D$. The distance depends on what
subspace of genes $G$ is used hence the predictor depends both on the training data
and $G$. Sometimes we will explicitly denote this dependence by writing the
predictor as $\p_G$.

We will use variants
of this basic predictor when constructing a scoring function that we discuss below.
For this we will not only need the closest point, but the values of the distances
to all sample points.

\subsection*{Evolution Algorithms}
\label{sub:evolution}

Starting off with an ensemble of different gene subspaces 
we want to determine
rules to evolve it to a new one that gives a better set of predictors.
To do this, we have to have a measure of how well a predictor
classifies samples into separate types.
We do this by means of a scoring function. 

\subsubsection*{Scoring function}
\label{sub:scoring}

The scoring function is used
to determine how well the predictor predicts data.
By definition we cannot use any information from the independent test data
in the development of the predictors. Therefore we consider only the training
data ${\cal D}_t$ to determine the fitness of a predictor. 
In other words, we need to score the effectiveness of the predictor
using only ${\cal D}_t$. This is done as follows. 
\begin{enumerate}
\item
We consider one
point $D_p$ in ${\cal D}_t$ as a pseudo test data point, and eliminate this point 
from ${\D}_t$, calling the resultant training data $\D_t'$. We then loop over all
points in $D_p \in \D_t$ in the following steps.
\begin{enumerate}
\item
We find the set of distances between 
$D_p$ and points in ${\D}_t'$. 
\item
If the type of the point giving the shortest distance matches the type
of $D_p$, we add $1$ to the scoring function. Otherwise we add nothing
and skip the remaining steps, continuing to loop over the remaining $D_p$'s.
\item
We consider the distances grouped by the classification type
of the target points. We consider the shortest distance of each type 
from $D_p$, whch we call $d_1,d_2,\dots, d_{N_T}$. 
\item
Of these we take the two shortest, $d_i$ and $d_j$ and add $C|d^2_i - d^2_j|$
where C is a constant chosen so that the value of this added term is $\ll 1$.
\end{enumerate}
\end{enumerate}

The scoring function depends on the predictor, which in turn is determined
by the training data and the subspace of genes $G$. We will denote this
latter dependence as $\s_G$

\subsubsection*{Initial Gene Pool}

Often it is necessary to narrow down the genes that are considered from the many
thousand that are measured on the microarray down to of order $10^2$ that are most
relevant. There are many ways of doing this, one common method being principle
component analysis. For the purposes here we choose instead a different method
that is highly effective.

We consider how genes distinguish two types $t_1,t_2 \in {\cal T}$ from each
other.
For each gene $g \in \G_t$ we consider its expression levels in the training
samples. We rank all the training samples in terms of the expression level of $g$.
We are looking for genes that for high levels give type $t_1$ and
for low levels give type $t_2$ (or vice-versa). 
When ranked this way, they sometimes will perfectly separate, that is
the first part of the list is one type, and the last part is the other. These
genes are ranked the highest. Most of the time however, a gene will not separate
so clearly and there will be overlapping regions.
Those with more overlaps of different
types are ranked lower. In this way we have a ranking of the genes that are
best able to distinguish $t_1$ from $t_2$, and we pick the top $M$ genes.

We then consider all distinct combinations of  $t_1$ and $t_2$ and pick
the best $M$ genes from each combination.  Genes may overlap, narrowing the
initial pool.  This is our initial set of
genes $\G_i$ that we will consider. 

\subsubsection*{Statistical Replication}

In analogy with statistical mechanics, we can think of the scoring
function as (negative) energy and invent a dynamics that evolves them
towards the highest scoring (lowest energy)  states. We do this at finite
temperature to allow the system to accept predictors that occasionally
may be less fit than their predecessors to get rid of local minima in
predictor space and to allow for a diverse population of predictors.

Suppose the system has evolved
to an ensemble of $n$ gene subspaces ${\cal E} \equiv \{G_1,G_2,\dots,G_n\}$, 
we will now employ a variant of a replication algorithm used in other
contexts\cite{ceperley} to replicate and modify each of the $G_i$'s.

\begin{enumerate}

\item For each $G \in {\cal E}$ we produce a new subspace as
follows.
\begin{enumerate}
\item A set of genes $G$ has genes $\{g_1,g_2,\dots,g_m\}$.
We randomly choose a gene $g_r$ from the  initial set $\G_i$, and add
it to $G$, producing a new set $G'$ of genes $\{g_1,g_2,\dots,g_m,g_r\}$.
If $g_r \in G$, $G' = G$. 
\item We compute the difference in the scoring functions 
$\delta\s ~=~ \s_{G'} -\s_{G}$.
\item We compute the weight for $G'$, $ w = \exp({\beta\delta\s})$,
where $\beta$ is the inverse ``temperature''. 
\end{enumerate}
\item Let $Z$ denote the sum  of these weights. We normalize the
weights by multiplying them by $n/Z$. 
\item We replicate all subspaces according to their weights.
With a weight $w$, the subspace is replicated $[w]$ and an
additional time with probability $w-[w]$. Here
$[w]$ denotes the largest integer $< ~ w$. 
\end{enumerate}

In summary, every subspace in the system is mutated and replicated
in accordance with how much fitter it was than its predecessor.
By carefully normalizing the system, the number of subspaces in the ensemble
stays close to $m$. Note that we can also do more than one potential mutation
in step 1. We will generalize this to allow $n_m$ potential mutations.

\subsubsection*{Annealing}

As the system evolves, the scoring function gives similar answers for all
members of the ensemble. In order to improve convergence, it is useful
to make the temperature a function of the spread in scores (or energy).
A variety of schedules for the temperature were tested.
We have found that good results are obtained if $\beta$ is taken 
to be $2/\Delta E$, where $\Delta E$ is the maximum spread in scores
between different members of the ensemble.

This is particularly useful because the scoring function has two basic
components. The first adds unity every time a sample is correctly
classified.  The second adds a much smaller number proportional to the
constant $C$ defined above, which is chosen to make this second component
$\ll 1$.  The second component tries to maximize the separation between
the different classes.  When the predictors have a evolved so that they
are all classifying correctly, we would like the second part to take
effect. By lowering the temperature by a schedule such as the one above,
the algorithm will then select for predictors that maximize the second
part of the scoring function. This leads to a much better set of genes.

\subsubsection*{Deterministic Evolution}

As an alternative to the statistical replication method above, we
also employed a method that is computationally more expensive but that
often performs 
better. The statistical method does not explore all possible combinations
of genes at each stage of growth. This can miss optimal gene combinations.
We get around this by a deterministic exploration of the optimum gene combinations
at every step. A single step goes as follows:
\begin{enumerate}
\item  Construct all distinct unions of  the $G$'s in the ensemble $\cal E$
with individual genes in the initial gene pool $\G_i$, i.e. 
$g_1,g_2,\dots,g_m,g_i$.
\item Sort all of these combinations by their score, keeping the
top $n_{top}$ of them.
\end{enumerate}

To save computer time we tried various values for $n_{top}$. It was found that
$n_{top} = n$, (the number of $G$'s in the ensemble) performed quite well. 
Another variant was to construct only half the unions and keep the top $n$,
for computational efficiency.

\section*{Results}

We now discuss application of the above algorithm to two data sets. The
work on the small
round blue cell tumors (SRBCT) of childhood and
the work on human acute leukemia\cite{Golub:leukemia}.

\subsection*{SRBCT Data}

Small round blue cell
tumors (SRBCT) of childhood are hard to classify by current clinical techniques. 
They appear similar under a light microscope and several techniques are normally
needed to obtain an accurate diagnosis. 
The paper \cite{Khan:srbct} used microarrays to study their classification
using of a single layer neural network. This work differed from previous
studies in that they were attempting to distinguish between four different
cancer types instead of the more usual 2.  They used 63 samples for training
and tested with 20.  By using a clever method combining principle component
analysis and sensitivity of their neural network to a gene, they were able
to reduce the number genes needed to 96 yet still classify all different forms
of cancer in test data perfectly.

Here we use the same data set to reduce the number of genes needed
and still classify the test data perfectly.

Starting with their data set of 2308 genes,
we constructed the initial pool of genes by considering
how well a gene discriminates type $i$ cancer from type $j$, as described above.
Since there
are 4 possible types, we have 6 combinations of $i$ and $j$. For
each of these we take the top 10 genes best able to discriminate for each $i,j$
pair. This gives a total of 50 genes, because it turns out that 10 of these
overlap between groups.

\begin{figure}[tb]
\centerline{
  \psfig{figure= 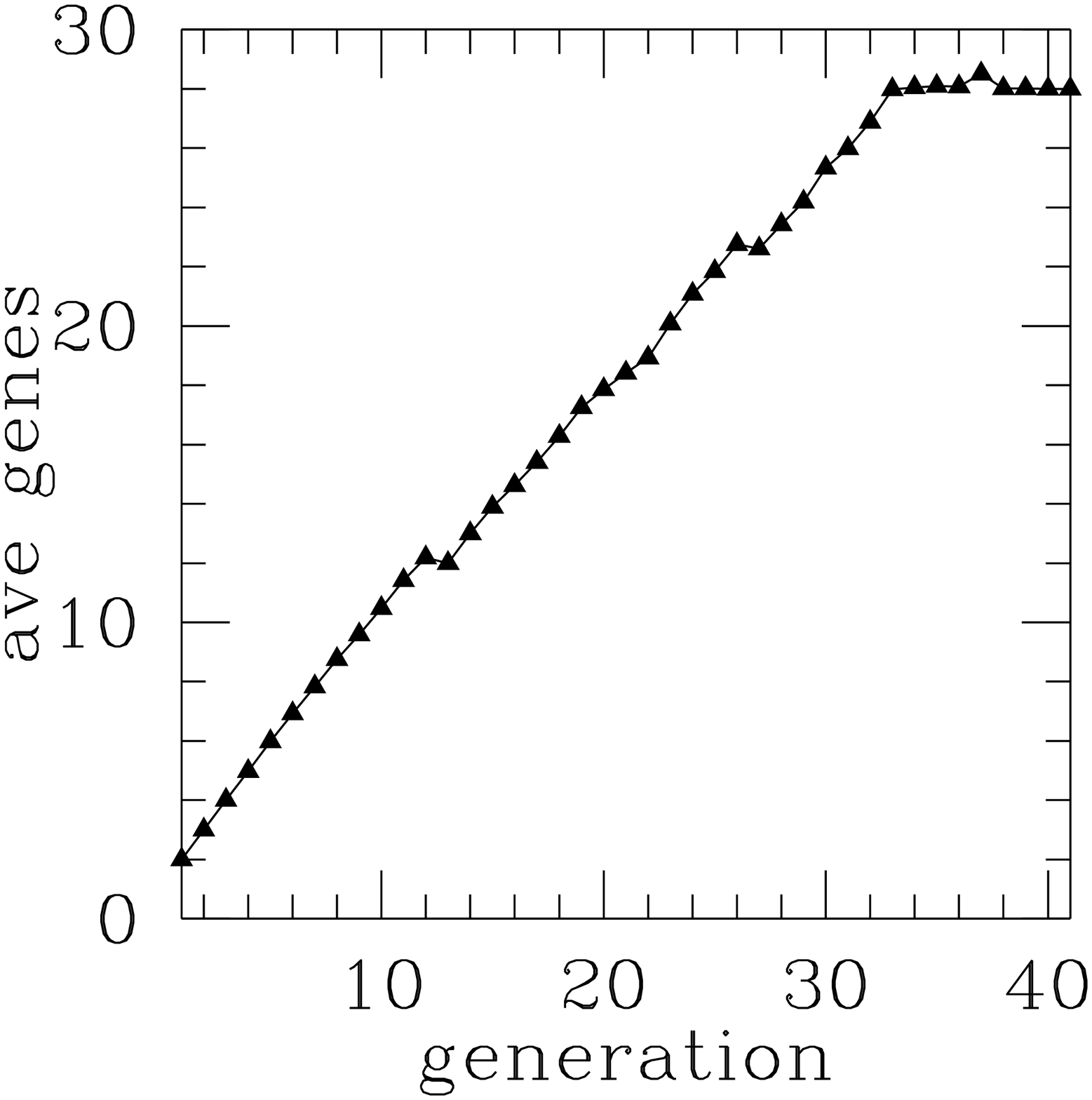,height=9.0cm}%width=12.0cm
}
\caption{\small{
The average number of genes for an ensemble of predictors
as a function of the number of generations,
for SRBCT data\cite{Khan:srbct}. The number of genes used here was
50, and the algorithm used was a statistical replication algorithm ($n_m = 2$).
}}
\label{fig:repl60a}
\end{figure}

\begin{figure}[tb]
\centerline{
  \psfig{figure=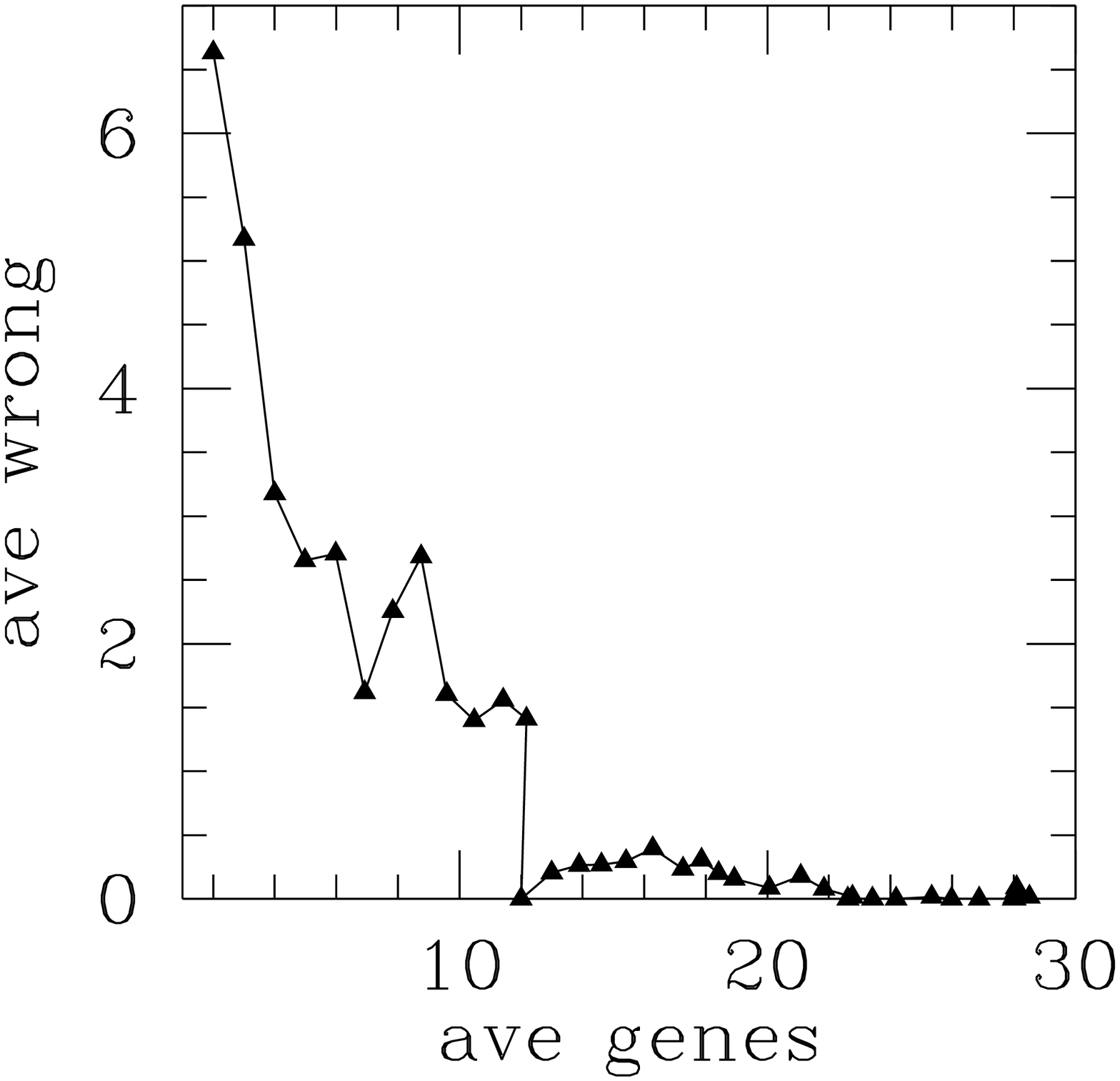,height=9.0cm}%width=12.0cm
}
\caption{\small{
The average number of mistakes made as a function of the
average number of genes in a predictor for the same parameters as in 
fig. \ref{fig:repl60a}.
}}
\label{fig:repl60b}
\end{figure}

We then evolve these gene subspaces according to the statistical replication method
described above.  Fig. \ref{fig:repl60a} shows the average number of genes in a predictor
as a function of the number of  generations.  The average is over the ensemble of predictors.
It starts to level off significantly at the 38th generation, because the addition
of further genes does not improve the scoring.
Fig.  \ref{fig:repl60b} shows how as a function of the average number of genes,
the predictors fair with the test data of 20 samples. The vertical axis is the 
average number of incorrect assignments, again averaged over all predictors
in the ensemble. By the 26th generation, more than 90\% of predictors perform
perfectly with the test data, and by the 41st generation, all predictors perform
perfectly using an average of 28 genes.

\begin{figure}[tb]
\centerline{
  \psfig{figure=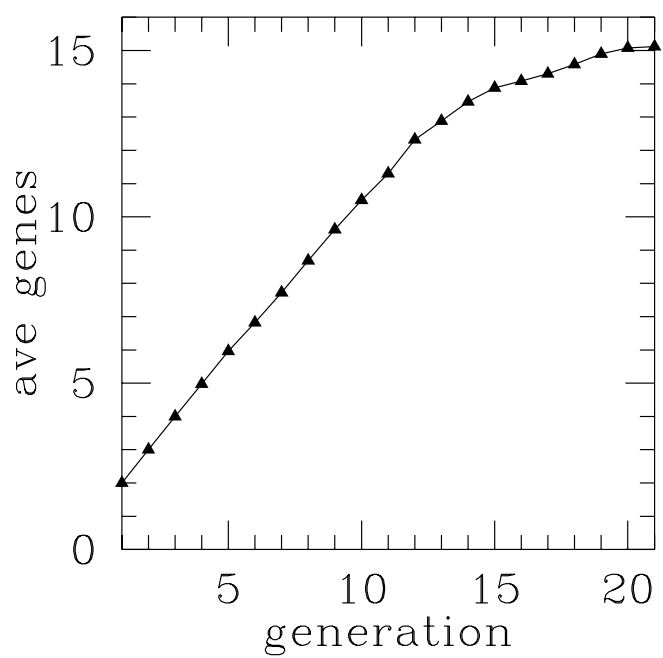,height=9.0cm}%width=12.0cm
}
\caption{\small{
The average number of genes for an ensemble of predictors
as a function of the number of generations for SRBCT data\cite{Khan:srbct}. The number of genes used here was
50, and the algorithm used was deterministic described in the text.
}}
\label{fig:det60a}
\end{figure}

\begin{figure}[tb]
\centerline{
  \psfig{figure=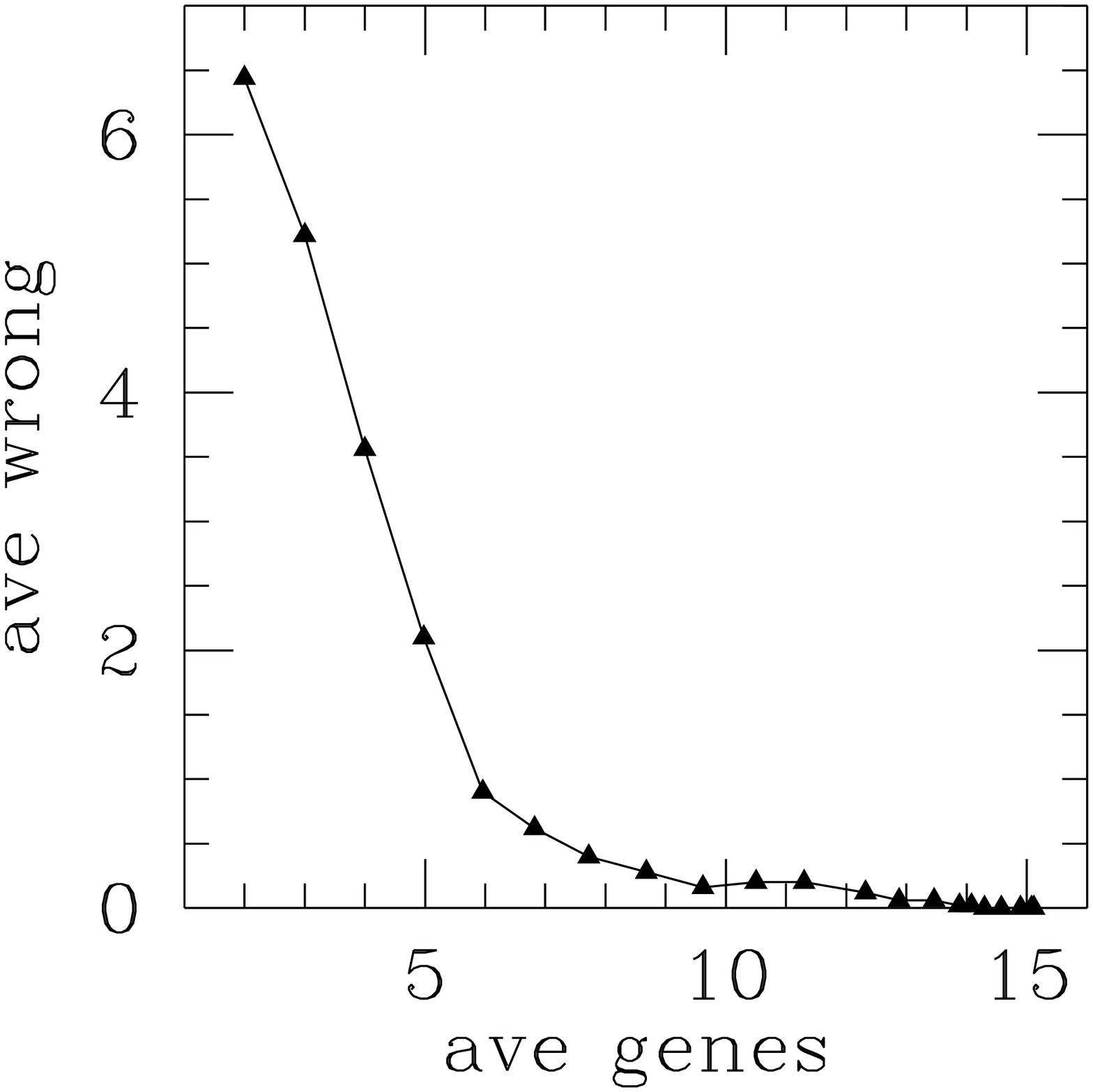,height=9.0cm}%width=12.0cm
}
\caption{\small{
The average number of mistakes made as a function of the
average number of genes in a predictor for the same parameters as in 
fig. \ref{fig:det60a}.
}}
\label{fig:det60b}
\end{figure}

We next use the deterministic evolution method described above only constructing
half the unions.
Fig.  \ref{fig:det60a} shows the average number of genes in a predictor
as a function of the number of  generations.  In this case it plateaus
off sooner, after about 15 generations. The comparison with the test data
is shown in Fig.  \ref{fig:det60b}. Here all predictors perform
perfectly when the average number of genes in a predictor is 15.1.
Here we only used half the possible unions and kept $n_{top} = 50$.

\begin{figure}[tb]
\centerline{
  \psfig{figure=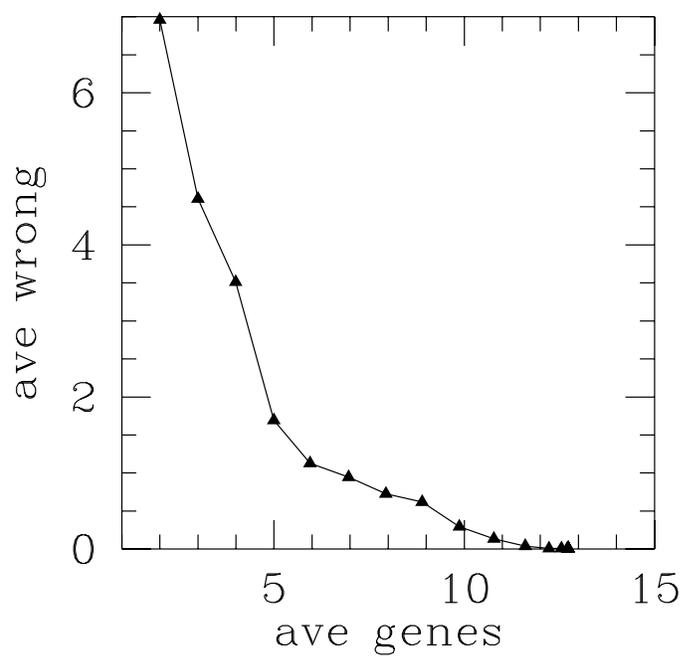,height=9.0cm}%width=12.0cm
}
\caption{\small{
The average number of mistakes made as a function of the
average number of genes in a predictor for an initial pool of 75 genes
SRBCT data\cite{Khan:srbct}. The parameters are described in the text.
}}
\label{fig:det90}
\end{figure}

Encouraged by the above results we did a larger run
starting with an initial pool of 90 genes of which 15 overlapped,
giving a total of 75 initial genes. Evolving these with $n_{top} = 150$ gives
the results shown in fig. \ref{fig:det90}. Of the top 100 predictors, all
predicted the test data perfectly. 
The average number of genes in a predictor was 12.7.

With this data, GESSES can be used to give an ensemble of predictors that have
perfect or near perfect performance. However if the initial gene pool is
reduced to below 60 genes, it degrades. 
For example, starting with 
the top 48 genes (giving 41 distinct genes)  with $n_{top} = 41 $ leads to a set of predictors that
make an average of 0.439 mistakes, and an average number of genes of 11.24. 
Despite this, one should keep in mind that over half the predictors predict the test data perfectly.
But starting with only
the top 24 genes (20 distinct genes)  with $n_{top} = 20$ leads to a set of predictors that
make an average of 1.45 mistakes, and an average number of genes of 10.95.

The genes found by these methods are mostly a subset of those found previously\cite{Khan:srbct}.
For example with 75 initial genes as described above (fig. \ref{fig:det90}), 
the union of all predictor genes
found in the top 100 predictors 
gave a total of 24 genes. These were a subset of the 96 Khan et. al. genes.
These are shown in Table \ref{table:dim90genes}, (excluding three genes that occur only once
among all the predictors).
However with the data of fig. \ref{fig:repl60b}, 
we find that out of a total of 25 different genes that comprise
all the possible genes used by the 50 predictors, four are different than those found
by Khan et. al. Of those four, one of them appear only once , and two of them 
occur quite frequently in the predictors. One, neurofibromin 2 appears in all predictors, and the other
thioredoxin appears in 37 of the 50 predictors. The third, homeobox B7 appears 6 times.
Neurofibromin has been associated with tumorigenesis\cite{reedguttmann:nf2}. 
It is believed that thioredoxin may play a role in cancer  and
Thioredoxin-1 is often associated with aggressive tumor growth\cite{powismontfort}.
In a study on multiple carcinigenesis of mouse skin\cite{changkozono}, Homeobox B7 appears to be
expressed at a much lower level than in normal mouse skin. Because this gene only appears
in 16\% of predictors, this may not be a significant correlation.

\subsection*{Leukemia Data}

Microarray data was obtained from patients having two types of leukemia, acute lymphoblastic leukemia (ALL), and
acute myeloid leukemia (AML). The data here was taken from bone marrow samples and the samples
were of different cell types, for example B or T cells and different patient genders.
Each sample was analyzed using  an Affymetrix microarrays containing expression levels
of 7129 genes. The data was divided into 38 training data points and 34 test points.

\begin{figure}[tb]
\centerline{
  \psfig{figure=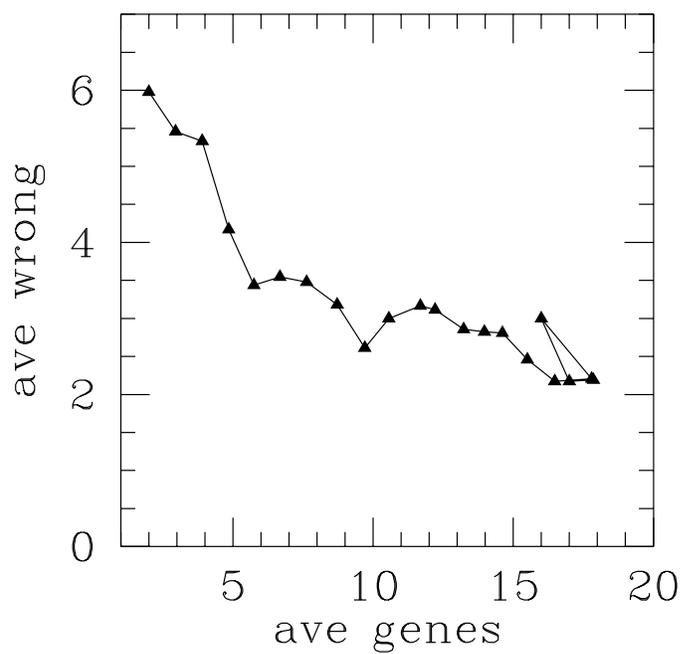,height=9.0cm}%width=12.0cm
}
\caption{\small{
The average number of mistakes for an ensemble of predictors
as a function of the average number of genes, for the leukemia data\cite{Golub:leukemia}. 
The number of genes used here was
50, and the algorithm used was a statistical replication algorithm ($n_m = 2$).
Note the curve is not singled valued because as the predictor evolves,
the average number of genes and number of mistakes can increase due to
statistical fluctuations.
}}
\label{fig:repl50}
\end{figure}

\begin{figure}[tb]
\centerline{
  \psfig{figure=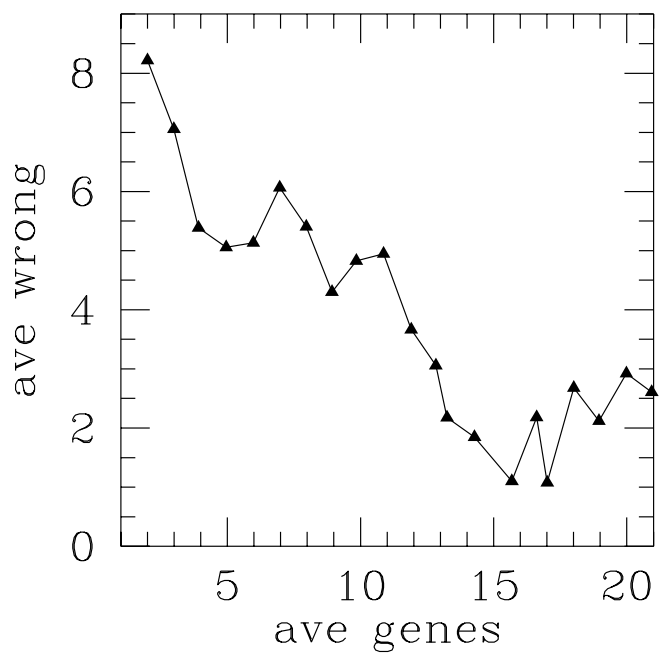,height=9.0cm}%width=12.0cm
}
\caption{\small{
The average number of mistakes for an ensemble of predictors
as a function of the average number of genes, for the leukemia data\cite{Golub:leukemia}. 
The number of genes used here was
200, and the algorithm used was a statistical replication algorithm ($n_m = 2$).
}}
\label{fig:repl200}
\end{figure}

Using the statistical replication algorithm we evolved the predictors and measured
the averaged number of misclassifications made as a function of the number of 
generations. This is done with an initial pool of 50 genes and the 
resulsts are shown in fig. (\ref{fig:repl50}).
The number of mistakes made by the ensemble of predictors plateaus at about 2.
The predictors vary in accuracy;
there are predictors that make no mistakes and some that make several. There appears
to be no way of distinguishing between them short of using the test data.
Data with 200 genes, fig. (\ref{fig:repl200}), shows a similar pattern but does not
completely plateau fluctuating in the average number of mistakes from about 1 to three.

\begin{figure}[tb]
\centerline{
  \psfig{figure=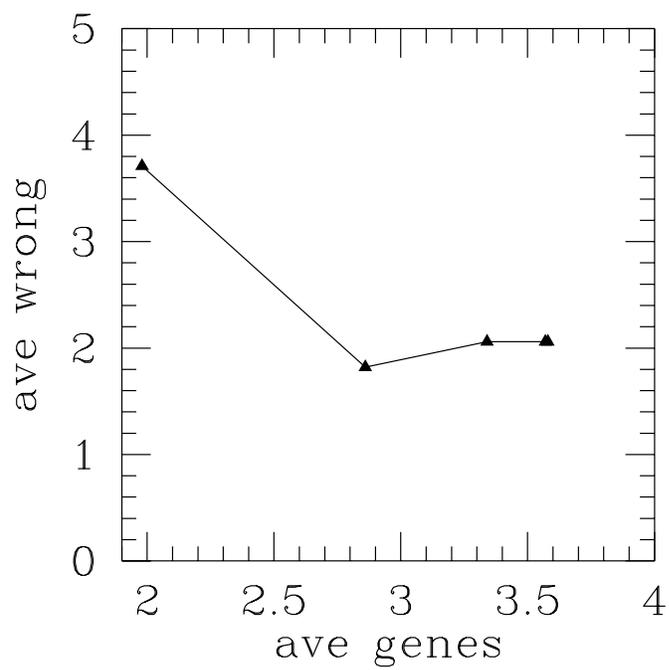,height=9.0cm}%width=12.0cm
}
\caption{\small{
The average number of mistakes for an ensemble of predictors
as a function of the average number of genes, for the leukemia data\cite{Golub:leukemia}. 
The number of genes used here was
200, and the algorithm used was a deterministic algorithm.
}}
\label{fig:det50}
\end{figure}

On the other hand using the deterministic evolution algorithm, we find a much faster
convergence to a steady state ensemble of predictors. Using an initial gene pool of
50 and $n_{top}$ of 100, the number of mistakes goes to about 2 with only three genes
in a predictor. This is shown in fig. (\ref{fig:det50}). The lack of convergence
to near perfect predictors is in agreement with other work on this 
data set\cite{Furey:svm2000,LiYang:mingenes,Golub:leukemia}.

Varying parameters such as the initial number of genes, $n_{top}$, and the method
of scoring does not lead to a statistically significant improvement in the 
average number of mistakes made. Also, as the above cases illustrate, 
the optimum number of genes in a predictors
varies between 3 to 25 depending on parameters. This is consistent with recent 
work on this data where also no clear cutoff in the number of genes needed for
an optimal predictor was also found\cite{LiYang:mingenes}. 

\section*{Discussion}

This paper has described a new and highly effective 
method, GESSES, that reduces the number of
genes necessary to perform an accurate classification. 
We implemented and tested it, producing an ensemble of  predictors
that use a minimal number of genes to perform a diagnosis of a cancer from
microarray data. 

There are many different kinds of prediction algorithms  that can be used
besides the nearest neighbor algorithm that we chose, 
among them are artificial neural networks, logistic regression, support vector machines
which appear to perform similarly. 

We have used a nearest neighbor search method for a variety of reasons.
It will classify training points perfectly. It makes little in the way of
assumptions of how new data extrapolates from old data. And in conjunction with
the replication algorithms used here it is quite efficient because 
it ``learns'' rapidly.
However which kind of predictor that is used is not the most important part of this work
and the replication algorithm could be implemented with anyone of the prediction
methods mentioned above. 

The main point is that evolutionary algorithms can be used to
determine minimal gene sets for tissue classification. By starting off with
an initial pool of candidate genes, an ensemble of predictors is evolved on
training data. Each predictor uses a different set of genes and its fitness
is scored by analyzing how well it separates the training data into 
separate classes. The system evolves converging to a set of predictors that
can be evaluated using test data.

In the case of SRBCT data\cite{Khan:srbct}, this method was able to find predictors using fewer than
15 genes that were able to reliably classify test data into one of four 
groups. Some of the genes found were different than the 96 found earlier\cite{Khan:srbct}
to do this classification and may be of biological significant. The optimum number of genes
to use in a predictor is approximately $12 \pm 2$.

In the case of leukemia data\cite{Golub:leukemia}, less useful
information can be obtained.  It is probably not possible to use the
training data to reliably construct a perfect predictor. It is clear
that more data is needed before the same level of prediction can be
achieved as with the SRBCT data.  This is in accord with other groups
findings\cite{Golub:leukemia,Furey:svm2000,LiYang:mingenes}.  At this
point it is not possible to come up with the optimal number of genes needed to
predict this data\cite{LiYang:mingenes}.  The main conclusion that one draws from this is that
there are many relevant genes in the diagnosis of cancers. However if
the data is not complete or is too noisy, it is not possible to exploit
this information to its full capacity.

It is hoped that using such a small set of genes could help lead to practical
uses of gene expression levels in cancer diagnosis, as it might turn out to be
more practical to build devices containing only 15 oligonucleotides rather than
thousands. It also might help to further the understanding of how the genes found
relate to the biology of these cancers.

\section*{Acknowledgments}

The author thanks Francoise Chanut for useful discussions.

\newpage

%\bibliographystyle{bioinformatics}
%\bibliography{microarray}

%{\small

%}

\newpage

\begin{table}
\begin{center}
\begin{tabular}{|l|l|}
\hline
id\# & gene description \\
\hline
365826 &  growth arrest-specific 1\\
298062 & "troponin T2, cardiac"\\
383188 &  recoverin\\
296448  & insulin-like growth factor 2 (somatomedin A)\\
769959 &  "collagen, type IV, alpha 2"\\
377461 &  "caveolin 1, caveolae protein, 22kD"\\
325182 &  "cadherin 2, N-cadherin (neuronal)"\\
1473131 &  "transducin-like enhancer of split 2, homolog of Drosophila E(sp1)"\\
207274 &  Human DNA for insulin-like growth factor II (IGF-2); exon 7 and additional ORF\\
357031 &  "tumor necrosis factor, alpha-induced protein 6"\\
812105 &  transmembrane protein\\
241412 &  E74-like factor 1 (ets domain transcription factor)\\
183337 &  "major histocompatibility complex, class II, DM alpha"\\
796258 &  "sarcoglycan, alpha (50kD dystrophin-associated glycoprotein)"\\
866702 &  "protein tyrosine phosphatase, non-receptor type 13 (APO-1/CD95 (Fas)-associated phosphatase)"\\
770394 &  "Fc fragment of IgG, receptor, transporter, alpha"\\
52076 &  olfactomedinrelated ER localized protein\\
609663 &  "protein kinase, cAMP-dependent, regulatory, type II, beta"\\
814260 &  follicular lymphoma variant translocation 1\\
784224 &  fibroblast growth factor receptor 4\\
295985 &  ESTs\\
\hline
\end{tabular}
\end{center}
\caption{
Genes found that perfectly predict SRBCT samples
}
\label{table:dim90genes}
\end{table}

\end{document}